\input amstex
\documentstyle{amsppt}
\magnification=1200
\hfuzz=25pt
%
%
\def\al{\alpha}

\def\de{\delta}    \def\De{\Delta}

%
%

\def\bI{{\bold I}}

\def\bS{{\bold S}}
\def\bV{{\bold V}}

%
%

\def\cI{{\Cal I}}

\def\cM{{\Cal M}}

\def\cO{{\Cal O}}
\def\cP{{\Cal P}}

%
 
%

\def\CC{{\Bbb C}}
\def\ZZ{{\Bbb Z}}

\def\PP{{\Bbb P}}

\def\eql{~=~}

\def\qeii{QEI$\!$I}
\def\wh#1{\widehat{#1}}
\def\deg{{\text{\rom{deg}}}}
\def\dim{{\text{\rom{dim}}}}

\def\eqv{~\equiv~}
\def\qn#1{(q)_{#1}}
\def\qbin#1#2{ \left[ \matrix {#1} \\ {#2} \endmatrix \right] }
\def\half{\textstyle{1\over2}}
\def\cM{{\Cal M}}
\def\shskip{\medskip}

\rightline{ADP-99-1/M76}
\rightline{\tt math/9902010}\bigskip\bigskip

\topmatter
\title $q$-identities and affinized projective varieties\\
I. Quadratic monomial ideals\endtitle
\rightheadtext{$q$-identities and affinized projective varieties, I}
\author Peter Bouwknegt \endauthor
\address Department of Physics and Mathematical Physics, University of 
Adelaide, Adelaide SA~5005, AUSTRALIA\endaddress
\email pbouwkne\@physics.adelaide.edu.au\endemail
\abstract 
We define the concept of an affinized projective variety and show
how one can, in principle, obtain $q$-identities by different 
ways of computing the Hilbert series of such a variety.
We carry out this program for projective varieties associated to
quadratic monomial ideals.
The resulting identities have applications
in describing systems of quasi-particles containing null-states and can
be interpreted as alternating sums of quasi-particle Fock space characters.
\endabstract
\endtopmatter
\document

\head 1. Introduction \endhead

The topic of $q$-identities, such as the Rogers-Ramanujan identities,
has attracted a lot of attention throughout the last century or so.
Initially, mostly in connection to the theory of partitions (see,
e.g., \cite{An}), later in connection with the representation theory
of infinite dimensional Lie algebras (see, e.g., \cite{Ka}).
Recently, there has been a surge of new research in this area
instigated by the discovery by the `Stony Brook group' of certain
`fermionic-type' (or quasi-particle type) formulas for the (chiral)
partition functions of two-dimensional conformal field theories (cf.,
in particular, the reviews \cite{DKKMM,KMM} and references therein).

There are many techniques for finding and/or proving $q$-identities
such as classical techniques by combinatorics, generating series, 
recursion relations as well as more modern ones based on
Bailey's transform, crystal bases, spinon bases and path representations.
The aim of this paper is to explain yet another technique based on the
relation between certain $q$-identities and the geometry of so-called
affinized projective varieties, in particular through the computation
of the Hilbert series of the (homogeneous) coordinate ring of such
varieties.  A relation between $q$-identities and the geometry of
infinite dimensional varieties has also been put forward in \cite{FS}.

One of the simplest examples of the type of identity we have in mind is
$$
{q^{M_1M_2} \over \qn{M_1} \qn{M_2} } \eql
\sum_{m\geq0} (-1)^m { q^{{1\over2}m(m-1)} \over \qn{m} \qn{M_1-m} 
  \qn{M_2-m}}  \,,
\tag{1.1}
$$
where 
$$
(q)_N \eql \prod_{k=1}^N (1-q^k)\,. \tag{1.2}
$$
The identity (1.1) first arose in the spinon description of
$\widehat{\frak{sl}_3}$ modules \cite{BS1,BS2}.  The alternating sum
on the right hand side of (1.1) indicates the presence of null-states
in the spinon Fock space which are removed by inclusion-exclusion
(sieving).  In this paper we will explain the geometric origin of the
identity (1.1) and an algorithm for constructing a host of identities
of similar (alternating) type which have similar interpretations as
alternating sums over quasi-particle Fock space characters.

To get some insight in the geometric origin of this relation, multiply
both sides by $y_1^{M_1} y_2^{M_2}$ and sum over $M_1,M_2\geq0$.  Then
consider the ${\Cal O}(q^0)$-term on each side of the equation.  On
the left hand side we have contributions from those $M_1,M_2\geq0$
such that $M_1M_2=0$, i.e., either $M_1\geq0$, $M_2=0$, or $M_1=0$,
$M_2\geq1$, while on the right hand side only the $m=0,1$ terms
contribute.  At ${\Cal O}(q^0)$ we thus find the (obvious) identity
$$
{1\over 1-y_1} + {y_2\over 1-y_2} \eql 
{1\over (1-y_1)(1-y_2)} - 
{y_1y_2 \over (1-y_1)(1-y_2)} \,.\tag{1.3}
$$
Alternatively, (1.3) arises from two different ways of computing the
Hilbert series of the projective variety $V$ consisting of 2 points in
$\PP^1$.  The left hand side of (1.3) is computed by constructing an
explicit basis for the homogeneous coordinate ring
$\CC[x_1,x_2]/\langle x_1x_2\rangle$ of $V$, while the right hand side
arises from a free resolution of this coordinate ring.  In this paper
we will argue that (1.1) arises, in a similar way, from an
appropriately defined affinization of the variety $V$.

The paper is organized as follows.  In section 2 we give a basic
review of some elementary concepts involving (projective) varieties,
Hilbert series and resolutions of monomial ideals in polynomial rings.
This section serves mainly to establish notations and to make the
paper accessible to an audience without expertise in algebraic
geometry.  In section 3 we introduce the concept of an affinized
projective variety and its associated Hilbert series and illustrate
their use in the example which leads to (1.1).  In section 4 we
explain an algorithm which leads to a $q$-identity for any projective
variety associated to a quadratic monomial ideal.  In section 5 we
illustrate the algorithm by explicitly going through some examples.
In section 6 we conclude with some remarks regarding the existence and
nature of $q$-identities associated to more general ideals.

In a sequel to this paper we will apply our 
ideas to the $q$-identities associated to flag varieties
and their connection to the representation theory of affine Lie algebras
and modified Hall-Littlewood polynomials \cite{BH}.

\head 2. Projective varieties and Hilbert series\endhead

\subhead 2.1. Varieties versus ideals \endsubhead\shskip

We begin by summarizing some elementary facts 
from algebraic geometry (see, e.g.,
\cite{Ha,Ei,CLO1,CLO2}).  
Throughout this paper we will work over the field $\CC$ of
complex numbers.

An affine variety $V\subset \CC^n$ is the zero locus of a set of 
polynomials, $f_1,\ldots,f_t$, in the coordinate ring, $\CC[x_1,\ldots,x_n]$,
of $\CC^n$, i.e.,
$$
V \eql \{ (x_1,\ldots,x_n)\in \CC^n \ : \ f_i(x_1,\ldots,x_n) \eql 0\
\text{for all}\ 1\leq i \leq t\}\,. \tag{2.1}
$$
There is a close correspondence between affine varieties $V\in \CC^n$
and ideals in the polynomial ring $\CC[x_1,\ldots,x_n]$.  Namely,
for $V\in \CC^n$, we can define an ideal
$$
\bI(V) \eql \{ f\in \CC[x_1,\ldots,x_n] \ : \ f(x_1,\ldots,x_n)=0 \ 
\text{for all}\ (x_1,\ldots,x_n) \in V\}\,, \tag{2.2}
$$
while, conversely, for an ideal $I\in \CC[x_1,\ldots,x_n]$, we can define 
the set 
$$
\bV(I) \eql \{ (x_1,\ldots,x_n)\in\CC^n \ : \ f(x_1,\ldots,x_n)=0\  
\text{for all}\  f\in I\} \,. \tag{2.3}
$$
That $\bV(I)$ is actually an affine variety is assured by Hilbert's 
basis theorem which states that every ideal $I\in \CC[x_1,\ldots,x_n]$
has a finite generating set.  
Clearly, for any affine variety $V\subset \CC^n$ we have 
$\bV ( \bI (V)) = V$.  The converse is not true however.  The 
composition $\bI \circ
\bV$ is neither injective nor surjective.  Precisely which $I$ appear
in the image of $\bI \circ \bV$ is settled by Hilbert's Nullstellensatz,
which states
$$
\bI ( \bV (I) ) \eql \sqrt{I} \eqv
\{ f\in  \CC[x_1,\ldots,x_n]\ : \ \exists\, r>0, \ f^r \in I \}\,.\tag{2.4}
$$
Thus, there exists a 1--1 correspondence between affine varieties 
$V\subset \CC^n$ and radical ideals $I\in \CC[x_1,\ldots,x_n]$, i.e.,
ideals for which $\sqrt I =I$.

Let $\PP^{n-1}=\PP(\CC^{n})$ denote the (complex) projective space.
We will use homogeneous coordinates $[x_1,\ldots,x_{n}]$ for
$\PP^{n-1}$.  Then, a projective variety $V\subset\PP^{n-1}$ is the
zero locus of a set of homogeneous polynomials $f_1,\ldots,f_t$ in the
homogeneous coordinate ring $\CC[x_1,\ldots,x_{n}]$ of $\PP^{n-1}$.
In analogy with the above, we now have a correspondence between projective
varieties and homogeneous ideals $I\subset \CC[x_1,\ldots,x_{n}]$
(see, e.g., \cite{CLO1, Chapter 8} for more details).


\subhead 2.2. Hilbert series \endsubhead\shskip 

Consider a homogeneous ideal $I\subset \bS = \CC[x_1,\ldots,x_n]$.
Let $\bS(V)=\bS/I$ denote the homogeneous coordinate
ring of the associated projective variety $\bV(I)$ and let $\bS(V)_M$
denote the vector space of homogeneous polynomials of degree $M$ in
$\bS(V)$.  The function
$$
h_V(M) \eql \dim\, \bS(V)_M\,,\tag{2.5}
$$
is called the (projective) Hilbert function of $V$.  One can prove
that there exists a polynomial $p_V(M)$ such that for $M$ sufficiently
large we have $p_V(M)=h_V(M)$.  The polynomial $p_V(M)$ contains
important information about the variety $V$, e.g., the degree of
$p_V(M)$ is the dimension of $V$.  In this paper we will also use the
Hilbert series $h_V(y)$ of $V$, i.e., generating series for $h_V(M)$
$$
h_V(y) \eql \sum_{M\geq0} \ h_V(M)\, y^M\,.\tag{2.6}
$$
By slight abuse of notation we will also denote by $h_V(y)$ the
Hilbert series of any $\bS$-module $V$.

For any $\bS$-module $M$,
let $M(a)$ denote the same module with the degree shifted by $a$.
Clearly, for $M\geq a$,
$$
\dim\, \bS(-a)_M \eql \dim\, \bS_{M-a}\eql \binom{M-a+n}{n}\,,\tag{2.7}
$$
so that
$$
h_{\bS(-a)}(y) \eql \sum_{M\geq0} \ \dim\, \bS(-a)_M \, y^M \eql
  {y^a \over (1-y)^n }\,.\tag{2.8}
$$
There are (at least) two methods to explicitly compute the Hilbert 
series of a projective variety $\bV(I)$.  The first is by 
constructing an explicit basis for the $\bS$-module $\bS(V) = \bS/I$
(see, e.g., \cite{CLO1, Chapter 9} for a recipe in the case of monomial
ideals).  The second is by means of a free resolution of the 
$\bS$-module $\bS(V)$.  The existence of a 
(finite length) free resolution of $\bS(V)$, i.e., an exact sequence
$$
0 @>>> F^{(\nu)} @>>> \ldots @>d_3>> F^{(2)} 
@>d_2>> F^{(1)} @>d_1>> F^{(0)} \cong \bS @>>> \bS(V) @>>> 
0\,, \tag{2.9}
$$
where each $F^{(i)} = \oplus_j \bS(-a^{(i)}_{j})$ for some set of 
positive integers $a^{(i)}_{j}$, is guaranteed by Hilbert's syzygy theorem.
Applying the Euler-Poincar\'e principle to the resolution (2.9) yields
$$
h_V(y) \eql \sum_{i\geq0}\ (-1)^i h_{F^{(i)}}(y) \eql
\sum_{i,j} \ (-1)^i\ {y^{a^{(i)}_j} \over (1-y)^n } \,, \tag{2.10}
$$
where we have used (2.8).

In case the ideal $I$ is homogenous in various subsets of coordinates 
separately, the quotient module $\bS/I$ carries a multi-degree $M=(M_1,
\ldots,M_s)$.  The above constructions then have an obvious multi-degree
generalization.

As an example consider the variety $V$ defined by the ideal
$I = \langle x_1x_2\rangle \subset \CC[x_1,x_2]$.  This variety 
consists of two points in $\PP^1$.  The coordinate ring $\bS(V)=
\CC[x_1,x_2]/\langle x_1x_2\rangle$ 
carries a bi-degree $\deg(x_1)=(1,0)$, $\deg(x_2)=(0,1)$.  Obviously,
$\bS(V)$ has a basis $\{x_1^m,\,m\geq0\} \cup \{ x_2^m,\,m>0\}$ so that
$$
h_V(y) \eql {1\over 1-y_1} + {y_2\over 1-y_2}\,. \tag{2.11}
$$
On the other hand, the resolution of $\bS(V)$ looks like
$$
0 @>>> \bS(-1,-1) @>d_1>> \bS @>>> \bS(V) @>>>  0 \,,\tag{2.12}
$$
where the map $d_1:\bS(-1,-1) \to \bS$ is defined as $P\mapsto (x_1x_2)P$.
So, the multi-degree generalization of (2.10) leads to 
$$
h_V(y) \eql { 1 \over (1-y_1)(1-y_2) } - 
{y_1y_2 \over (1-y_1)(1-y_2) }\,. \tag{2.13}
$$
The equality of (2.11) and (2.13) is obvious (cf.\ (1.3)).

\subhead 2.3.  Taylor's resolution of a monomial ideal \endsubhead\shskip

In this section we recall a resolution of monomial ideals due 
to Taylor (see \cite{Ei, Exercise 17.11}).
Suppose we have an ideal $\langle f_1,\ldots,f_t\rangle\subset
\CC[x_1,\ldots,x_n]\equiv \bS$ generated by monomials $f_i$, $i=1,\ldots,t$.
Let $\cI_s$ be the set of (ordered) subsets of $\{1,\ldots,t\}$ of
length $s$, i.e., $I\in\cI_s$ is an $s$-tuple $\{i_1,\ldots,i_s\}\subset
\{1,2,\ldots,t\}$ with $i_1<\ldots<i_s$.  We will also denote
$\cI = \bigcup_s \cI_s$ and $|I|=s$ for $I\in\cI_s$.  Let $F^{(s)}$ be
the free $\bS$-module on basis elements $e_I, I\in\cI_s$, and, for
$I\in\cI_s$, let
$$
f_I \eql \text{LCM}\{f_i,\ i\in I\}\,, \tag{2.14}
$$
where LCM$\{f_i\}$ denotes the lowest common multiple of the monomials
$f_i$, $i\in I$.
Furthermore, for $I=\{i_1,\ldots,i_s\}
\in\cI_s$ and $J\in\cI_{s-1}$ we define
$$
c_{IJ} \eql \cases 0 & \text{if}\ J\not\subset I\,, \\
(-1)^k\, f_I/f_J & \text{if}\  I=J\cup \{i_k\}\ \text{for some}\ k\,.\endcases
\tag{2.15}
$$
We have maps $d_s: F^{(s)} \to F^{(s-1)}$ defined by
$$
d_s\ :\ e_I ~\mapsto~ \sum_{J\in\cI_{s-1}} \ c_{IJ}\, e_J\,, \tag{2.16}
$$
satisfying $d_{s-1}d_s=0$.  The corresponding complex
$$
0 @>>> F^{(t)} @>d_t>> \ldots  @>d_2>> F^{(1)}   
@>d_1>> F^{(0)} \cong \bS @>>> \bS(V) @>>> 0 \,, \tag{2.17}
$$
gives a free resolution of $\bS(V)$.

\remark{Remark 2.1} If $f_{i_1,\ldots,i_s} = f_{i_1} \ldots f_{i_s}$ for
all $I = \{ i_1,\ldots,i_s\}\in\cI$, the resolution (2.17) is a so-called
Koszul resolution and the corresponding variety $\bV(I)$ is called 
a complete intersection (cf.\ \cite{Ha, Example 13.16}).  In general,
there may exist subsets $J,J'\in\cI$ such that $f_I = f_J f_{J'}$ and
$J\cup J'\subset I$.  We will refer to these as `Koszul parts' of 
Taylor's resolution.  They are usually an indication that the resolution 
(2.17) is not a minimal resolution.
\endremark\medskip

As an example, consider the ideal $I=\langle x_1x_2,x_2x_3,x_3x_4\rangle
\subset \CC[x_1,x_2,x_3,x_4]$ (cf.\ section 5.2). 
Put $f_1=x_1x_2$, $f_2=x_2x_3$ and $f_3=x_3x_4$.
We find
$$\align
f_{12}  \eql x_1x_2x_3\,,\quad f_{23} & \eql x_2x_3x_4 \,,\quad f_{13} \eql 
x_1x_2x_3x_4 \,,\\
f_{123} & \eql x_1x_2x_3x_4\,, \tag{2.18}\endalign
$$
so Taylor's resolutions (2.17) is given by
$$ \align
F^{(1)} & ~\cong~ \wh\bS e_1 \oplus \wh\bS e_2 \oplus \wh\bS e_3 \\
& ~\cong~ \wh\bS(-1,-1,0,0) \oplus \wh\bS(0,-1,-1,0)\oplus \wh\bS(0,0,-1,-1)
  \,, \\
F^{(2)} & ~\cong~ \wh\bS e_{12} \oplus \wh\bS e_{23} \oplus \wh\bS e_{13} \\
& ~\cong~ \wh\bS(-1,-1,-1,0)\oplus \wh\bS(0,-1,-1,-1)
   \oplus\wh\bS(-1,-1,-1,-1)\,, \\
F^{(3)} & ~\cong~ \wh\bS e_{123} \\ & ~\cong~ \wh\bS(-1,-1,-1,-1) \,,
\tag{2.19}\endalign
$$
and maps $d_s$ given by (2.16).  The minimal resultion is however
given by removing the spaces $\wh\bS(-1,-1,-1,-1)$ from $F^{(2)}$ and
$F^{(3)}$, as one can easily see.

\head 3. Affinized projective varieties and $q$-identities\endhead

Consider a projective variety $V\subset
\PP^{n-1}$, defined by the ideal $\bI(V)$ generated by a set of 
homogeneous elements $f_i$, $i=1,\ldots,t$. By the affinized
projective variety $\widehat V\subset
\widehat{\PP^{n-1}}$ we mean the
infinite projective variety defined by the ideal $\bI(\widehat{V})$
generated by the relations $f_i[m]$, $i=1,\ldots,t$,
$m\in\ZZ_{\geq0}$, in $\wh{\bS} = \CC[x_1,\ldots,x_n]\otimes
\CC[\![t]\!]=
\CC[x_1[m],\ldots,x_n[m]]_{m\in\ZZ_{\geq0}}$ where we have 
written $x_i[m] = x_i \otimes t^m$, and where $f_i[m]$ is obtained
from $f_i$ by replacing all monomials $x_{i_1}\ldots x_{i_r}$ by
$$
(x_{i_1}\ldots x_{i_r})[m] \eql \sum_{n_1,\ldots,n_r\geq0\atop
n_1+\ldots +n_r=m} 
x_{i_1}[n_1]\ldots x_{i_r}[n_r]\,. \tag{3.1}
$$
The coordinate ring $\bS(\wh{V})$ of the affinized projective 
variety $\wh{V}$ is graded by the multi-degree defined by 
$$
\deg( x_i[m] ) \eql (\deg(x_i);m) \,,
\tag{3.2}
$$
i.e., both by the (multi-) degree 
inherited from the underlying projective variety $V$, as 
well as the `energy' $m$.

We denote by $\bS(\wh{V})_{(M;N)}$ the vector space of homogeneous 
polynomials $f$ of multi-degree $(M;N)$ in $\bS(\wh V)$.
By analogy with (2.5), the Hilbert function is defined as 
$h_{\wh V}(M;N) = \dim\,\bS(\wh{V})_{(M;N)}$.  Note that the 
introduction of `energy' makes $h_{\wh V}(M;N)$ finite.
The Hilbert series of $\wh{V}$ is defined as
$$
h_{\wh V}(y;q) \eql \sum_{M,N} \  h_{\wh V}(M;N)\ y^M\, q^N\,.
\tag{3.3}
$$
We will also be using the partial Hilbert series
$$
h_{\wh V}(M;q) \eql \sum_{N} \  h_{\wh V}(M;N)\ q^N\,.
\tag{3.4}
$$

\remark{Remark 3.1} 
Note that $\bS(V)\subset \bS(\wh V)$ through the identification
$x_i \sim x_i[0]$.  Therefore we have the obvious equality
$h_{\wh V}(M;0) = h_V(M)$ between the Hilbert function of $V$ and
the energy $N=0$ Hilbert function of $\wh V$, i.e., the 
$\cO(q^0)$-term in the partial Hilbert series of $\wh V$.
\endremark\medskip

\remark{Remark 3.2} The variables $x_i[m]$ combine into
`currents'
$$
x_i(t) \eql \sum_{m\geq0}\ x_i[m]\,t^m\,.
$$
In terms of these currents, the `energy' is just the eigenvalue of 
the derivation $t {d\over dt}$ while the ideal $\wh I$ is generated by 
the modes of currents $f_i(t)$ which are compositions of the $x_i(t)$.
The Hilbert series (3.4) has the interpretation of a $U(1)$ character.
\endremark\medskip

As in the finite dimensional case, there are in principle two 
different ways of computing the Hilbert series of $\wh V$.  On
the one hand, we may be able to construct an explicit basis for the
coordinate ring $\bS(\wh V)$ of $\wh V$.  On the other hand we may 
compute $h_{\wh V}(y;q)$ by applying the Eurler-Poincar\'e principle
to a free resolution of $\bS(\wh V)$
$$
\ldots @>d_3>> F^{(2)} @>d_2>> F^{(1)} @>d_1>> \wh{\bS} @>>> \bS(\wh{V})
@>>> 0\,,\tag{3.5}
$$
respecting the grading by the multi-degree (3.2).  
Of course, in the affinized case, the resolution (3.5) will be infinite,
but at every degree $(M;N)$ only a finite number of spaces contribute.

Comparing the results of the two different computations of $h_{\wh V}(y;q)$
will produce the required $q$-identity.

In section 3.2 we apply this idea to the example discussed in section
2.2 and show that this leads to the identity (1.1) alluded to in the
introduction.  Other examples based on quadratic monomial ideals will
be discussed in sections 4 and 5 of this paper.

In other cases we might have additional information on
$\bS(\wh{V})$, e.g., it can be that $\bS(\wh{V})$ admits the action of
a (Lie) algebra, in which case $h_{\wh V}(y;q)$ can actually be 
interpreted
as a character of this algebra, which might be known independently.
This will be a particularly useful point of view in the case of flag
varieties and will be the subject of a future publication \cite{BH}.

\subhead 3.2.  Prime example: $I= \langle x_1x_2\rangle$ \endsubhead\shskip

Consider again the projective variety $V$ defined by the ideal $I =\langle
x_1x_2\rangle \subset \CC[x_1,x_2]$ (cf.\ section 2.2).
The affinized variety $\wh V$ is defined by the ideal $\wh I\subset
\CC[x_1[m],x_2[m]] = \wh \bS$ 
generated by all $f[m], m\in\ZZ_{\geq0}$, where 
$$
f[m] \eql (x_1x_2)[m] \eql \sum_{r+s=m} \ x_1[r] x_2[s]\,.
\tag{3.6}
$$
We have a multi-degree on $\bS(\wh V) = \wh \bS/ \wh I$ defined by
$$
\deg(x_1[m]) \eql (1,0;m)\,,\qquad \deg(x_2[m]) \eql (0,1;m)\,.
\tag{3.7}
$$
Using the relations (3.6),  
it can be shown that a basis for ${\bS}(\wh{V})_{(M_1,M_2)}$ is given by
$$ \align
x_1[n^{(1)}_{M_1}] \ldots   x_1[n^{(1)}_1] & \, x_2[n^{(2)}_{M_2}]
\ldots x_2[n^{(2)}_1] \\
\text{with}\quad n^{(1)}_{M_1}\geq \ldots \geq n^{(1)}_1 \geq M_2 & \quad 
\text{and} \quad
n^{(2)}_{M_2} \geq \ldots \geq n^{(2)}_1\geq0 \,.
\tag{3.8}\endalign
$$

Before we prove (3.8), let us notice that by using 
$$
\sum_{n_1\geq \ldots\geq n_m\geq 0} q^{n_1+\ldots+n_m} 
 \eql {1\over \qn{m}} \,,
\tag{3.9}
$$
we immediately find 
$$
h_{\wh{V}}(M_1,M_2;q) \eql {q^{M_1M_2} \over 
 (q)_{M_1} (q)_{M_2} }\,.
\tag{3.10}
$$

\demo{Proof of (3.8)}
To prove the claim, we first have to show that every monomial 
$$
x_1[n^{(1)}_{M_1}] \ldots x_1[n^{(1)}_1] \, x_2[n^{(2)}_{M_2}]
\ldots x_2[n^{(2)}_1] \,, \tag{3.11}
$$
with $n^{(1)}_{M_1}\geq \ldots \geq n^{(1)}_1 \geq 0$ and
$n^{(2)}_{M_2} \geq \ldots \geq n^{(2)}_1\geq0$, can be written as a
linear combination of monomials (3.8) modulo terms in the ideal 
$\wh I$ generated by the $f[m]$.  Clearly, it suffices to prove this for
$M_1=1$ and $n^{(1)}_1=M_2-1$.  First, we claim that
$$
x_1[k] x_2[0]^{M_2} ~\in~ \wh I\,,\qquad \forall k\leq M_2-1,\, M_2\geq1\,.
\tag{3.12}
$$
This is proved by induction to $M_2$.
Obviously, for $M_2=1$, $x_1[0]x_2[0] = f[0]\in \wh I$.  The induction step
$M_2 \to M_2+1$ follows from
$$
x_1[M_2] x_2[0]^{M_2+1} ~\sim~ -\sum_{k=1}^{M_2} x_1[M_2-k] 
x_2[k] x_2[0]^{M_2-1} ~\in~ \wh I\,,
$$
where in the last step we have used the induction hypothesis and by
$\sim$ we denote equivalence upto terms in the ideal $\wh I$.  Now,
let $d$ denote the sum of the $(M_2-1)$-st smallest arguments of the
$x_2$-variables in the monomial (3.11), i.e., $d=\sum_{j=1}^{M_2-1}
n^{(2)}_j$.  We will prove, by a nested induction to $(M_2,d)$, that
each monomial
$$
x_1[M_2-1] x_2[n^{(2)}_{M_2}]\ldots x_2[n^{(2)}_1]\,,
$$
with $n^{(2)}_{M_2} \geq \ldots \geq n^{(2)}_1\geq0$, can be written 
in the form (3.8) modulo terms in the ideal.  Denote by $\cM$ the span of 
(3.8).  For $d=0$ and 
arbitrary $M_2\geq1$ we have, using (3.12),
$$\align
x_1[M_2-1] x_2[m] & x_2[0]^{M_2-1}  ~\sim~ 
- \sum_{k=0,\ldots, M_2-1+m\atop k\neq m} 
x_1[M_2-1+m-k] x_2[k]x_2[0]^{M_2-1}\\
 ~\sim~ & - \sum_{k=0,\ldots, m-1} 
x_1[M_2-1+m-k] x_2[k]x_2[0]^{M_2-1} \in \cM \,,\endalign
$$
where $m\in\ZZ_{\geq0}$ is arbitrary.  Now, for the induction step
$(M_2,d) \to (M_2,d+1)$,
assume the statement is true for all $M_2'=M_2, d'\leq d$ and $M_2'<M_2$,
all $d'$.  Consider
$$
x_1[M_2-1] x_2[n^{(2)}_{M_2}]\ldots x_2[n^{(2)}_1] \,,
$$
with $n^{(2)}_{M_2} \geq \ldots \geq n^{(2)}_1\geq0$ and
$\sum_{j=1}^{M_2-1} n^{(2)}_j=d+1$.  Omitting terms in $\cM$ we have 
$$ 
x_1[M_2-1] x_2[n^{(2)}_{M_2}]\ldots x_2[n^{(2)}_1] ~\sim~ 
\sum_{k=1}^{M_2-1} x_1[M_2-1-k] x_2[n^{(2)}_{M_2}+k]
x_2[n^{(2)}_{M_2-1}]\ldots x_2[n^{(2)}_1] \,.
$$
By the induction step we can write 
$$ \align
x_1[M_2-1-k] & x_2[n^{(2)}_{M_2}+k]
x_2[n^{(2)}_{M_2-1}]\ldots x_2[n^{(2)}_1] \\ ~\sim~ &
\sum_{m_1\leq\ldots \leq m_{M_2-1}} a^{(k)}_{m_1\ldots m_{M_2-1}}
x_1[M_2-1] x_2[n^{(2)}_{M_2}+k]x_2[m_{M_2-1}] \ldots x_2[m_1]\,, \endalign
$$
again modulo terms in $\cM$.  But now notice that in 
$$ 
\sum_{k=1}^{M_2-1} 
\sum_{m_1\leq\ldots \leq m_{M_2-1}} a^{(k)}_{m_1\ldots m_{M_2-1}}
x_1[M_2-1] x_2[n^{(2)}_{M_2}+k] x_2[m_{M_2-1}] \ldots x_2[m_1]\,,
$$
all terms have $d' \leq d$ and hence are in $\cM$ (modulo $\wh I$) by the
induction hypothesis.  The proof is completed once we show the converse 
statement, i.e., that no linear combination of monomials (3.8) is in the
ideal $\wh I$.  This is proved similarly as above. \qed
\enddemo

\remark{Remark 3.3} As an aside, we remark that
the monomials (3.8) are the complement of the leading terms of 
a Gr\"obner basis for $I$ with respect to the lexicographic order
defined by $x_1[0]>x_1[1]>\ldots >x_2[0]>x_2[1]>\ldots$. 
The Gr\"obner basis can in principle be found by applying 
Buchberger's algorithm (see, e.g., \cite{CLO1,CLO2}).
\endremark\medskip

On the other hand, a resolution of $\bS(\wh V)$ is easily constructed 
since in this case $\wh V$ is a complete intersection.  To this end, let
$\cI_m$ be the set of ordered $m$-tuples $\{n_1,\ldots,n_m\}$ with
$n_1> \ldots > n_m\geq0$, and let $F^{(m)}$ be the free $\wh\bS$-module
on generators $e_I$, $I\in\cI_m$, with $\deg(e_I) = (-m,-m;-n_1-\ldots
-n_m)$.  I.e., 
$$ 
F^{(1)}  ~\cong~ \bigoplus_{n\geq0}\ \wh\bS(-1,-1;-n) \,,\qquad
F^{(m)}  ~\cong~ \bigwedge{}^{\!m} \, F^{(1)} \,.\tag{3.13}
$$
For $I=\{n_1,\ldots,n_m\} \in\cI_m$, and $J\in \cI_{m-1}$ define 
$$
c_{IJ} \eql \cases 0 & \text{if}\ J\not\subset I\,, \\
  (-1)^k f[n_k] & \text{if} \ I=J\cup \{n_k\} \,.\endcases \tag{3.14}
$$
We have maps $d_m:F^{(m)} \to F^{(m-1)}$, satisfying 
$d_{m-1} d_m =0$,  defined by
$$
d_m\ : \ e_I ~\mapsto~ \sum_{J\in\cI_{m-1}} \ c_{IJ}\, e_J\,, \tag{3.15}
$$
and such that the complex
$$
\ldots  @>d_3 >> F^{(2)} @>d_2>> F^{(1)} @>d_1>> \wh\bS @> >> \bS(\wh V) 
@> >> 0 \,,\tag{3.16}
$$
is exact, i.e., provides a resolution of $\bS(\wh V)$.
Now, clearly,
$$ \align
h_{F^{(m)}}(M_1,M_2;q) & \eql \sum_{n_1>\ldots  >n_m \geq0}\ \sum_{N\geq0} 
 \text{dim}\, \bS(-m,-m;-n_1-\ldots -n_m)_{(M_1,M_2;N)}\, q^N \\
& \eql \sum_{n_1>\ldots >n_m \geq0}\ \sum_{N\geq0}  
 \text{dim}\, \bS_{(M_1-m,M_2-m;N-n_1-\ldots -n_m)}\, q^N \\
& \eql \sum_{n_1>\ldots >n_m \geq0} q^{n_1+\ldots+n_m} \ 
  \sum_{N\geq0} \text{dim}\, \bS_{(M_1-m,M_2-m;N)} q^N \\
& \eql 
{q^{ {1\over2}m(m-1) } \over (q)_m} {1\over (q)_{M_1-m} (q)_{M_2-m} }\,,
\tag{3.17}
\endalign
$$
where we have used (3.9).
Thus, by applying the Euler-Poincar\'e principle to the resolution (3.16),
we find
$$
h_{\wh{V}}(M_1,M_2;q) \eql \sum_{m\geq0} \ (-1)^m
{q^{ {1\over2}m(m-1) } \over (q)_m} {1\over (q)_{M_1-m} (q)_{M_2-m} } \,.
\tag{3.18}
$$

Equating the expressions (3.10) and (3.18) leads to the $q$-identity (1.1).
This identity first occurred in \cite{BS1,BS2} where it was 
used to compare two different 
quasi-particle descriptions of the $(\wh{\frak{sl}_3})_1$ modules
(see also \cite{BH}).  It was proved in \cite{BS2} by 
generating function techniques. 

The (full) Hilbert series (3.3) follows straightforwardly from either 
(3.10) or (3.18)
$$
h_{\wh V}(y;q) \eql {(y_1y_2;q)_\infty \over (y_1;q)_\infty (y_2;q)_\infty}\,,
\tag{3.19}
$$
where 
$$
(y;q)_N \eql \prod_{k=1}^N \ (1-y q^{k-1} ) \,.\tag{3.20}
$$

\remark{Remark 3.4} Instead of considering the affinized projective variety
$\wh V$, one may also consider a partial affinization $V_N$ defined 
as the (finite dimensional) projective variety associated to the 
ideal $I_N \subset \bS_N$,
where $\bS_N = \CC[ x_1[m],x_2[m] ]_{0\leq m\leq N}$ and 
$I_N = \langle f[m] \rangle _{0\leq m\leq N}$.
While an explicit monomial basis of $\bS(V_N) = \bS_N/I_N$, analogous to 
(3.8), is considerably more complicated than in the fully affinized case, 
the resolution of $\bS(V_N)$ is simply the restriction of the resolution 
(3.16) to all $m$-tuples $\{n_1,\ldots,n_m\}$ satisfying 
$N\geq n_1 > \ldots > n_m \geq0$.  Using the analogue of (3.9)
$$
\sum_{N\geq n_1\geq \ldots\geq n_m\geq 0} q^{n_1+\ldots+n_m} 
 \eql \qbin{N+m}{m} \,,
\tag{3.21}
$$
where 
$$
\qbin{m}{n} \eql { \qn{m} \over \qn{n} \qn{m-n} }\,,
\tag{3.22}
$$
denotes the Gaussian polynomial, we find the Hilbert series
$$
h_{V_N}(M_1,M_2;q) \eql \sum_{m\geq0} \ (-1)^m q^{ {1\over2} m(m-1)}
\qbin{N+1}{m} \qbin{N+M_1-m}{M_1-m} \qbin{N+M_2-m}{M_2-m} \,,
\tag{3.23}
$$
or, equivalently,
$$
h_{V_N}(y;q) \eql {(y_1y_2;q)_{N+1} \over (y_1;q)_{N+1} (y_2;q)_{N+1}}\,.
\tag{3.24}
$$
\endremark

\head 4. Quadratic monomial ideals \endhead

In this section we illustrate the procedure outlined 
in section 3 by discussing some examples of $q$-identities 
associated to projective varieties with monomial
quadratic defining relations.  

\subhead 4.1.  A basis of $\bS(\wh V)$ \endsubhead\shskip

Let $\cP$ be a set of (ordered) pairs $(i,j)$, $i<j$, with 
$i,j\in \{1,\ldots,n\}$.  Consider the quadratic monomial ideal
$I = \langle x_ix_j\rangle_{(i,j)\in\cP} \subset
\CC[x_1,\ldots,x_n] \equiv \bS$ with associated projective variety
$V=\bV(I)$.

We have a multi-degree $M=(M_1,\ldots,
M_n)$ on $\bS(V) = \CC[x_1,\ldots,x_n]/I$, where $M_i$ is the number 
of $x_i$ in a monomial $x^\al$.  Let $\wh{V}$ be the affinization of 
$V$ as defined in section 3.  
\proclaim{Theorem} A basis of $\bS(\wh{V})$ is given by the following 
monomials 
$$
x_1[n^{(1)}_{M_1}]\ldots x_1[n^{(1)}_1]x_2[n^{(2)}_{M_2}]\ldots
  x_2[n^{(2)}_{1}] \ldots x_n[n^{(n)}_{M_n}] \ldots x_n[n^{(n)}_{1}]\,,
\tag{4.1}
$$
with 
$$
n^{(i)}_{M_i}\geq \ldots \geq n^{(i)}_2\geq n^{(i)}_{1} \geq 
\sum_{j\atop (i,j)\in {\cP}} M_j\,,\qquad \forall i\,.\tag{4.2}
$$
\endproclaim

We will omit the proof, which is a straightforward generalization of 
the proof in section 3.2.

\remark{Remark 4.1} Note that the basis of $\bS(V)\subset \bS(\wh V)$
under the identification $x_i \sim x_i[0]$ induced by (4.1) is the 
obvious one.  A monomial $x_{i_1}\ldots x_{i_t} \in \bS(V)$ iff 
there is no pair $(i_r,i_s)$ such that $(i_r,i_s)\in \cP$.
\endremark\medskip

Using the basis (4.1), it immediately follows that the (partial)
Hilbert series of $\wh{V}$ is given by (cf.\ (3.10))
$$
h_{\wh{V}}(M_1,\ldots,M_n;q) \eql 
{q^{\sum_{(i,j)\in {\cP}} M_iM_j } \over \qn{M_1} \ldots \qn{M_n} } \,.
\tag{4.3}
$$

\subhead 4.2.  $q$-identities; the algorithm \endsubhead\shskip

Having found the Hilbert series of a general affinized projective
variety $\wh V$ corresponding to the affinization $\wh I$ of a
quadratic monomial ideal $I$, we can now, in principle, obtain a
$q$-identity by explicitly constructing a free resolution of the
coordinate ring $\bS(\wh V)$ of $\wh V$ as we have done in the example
of section 3.2.  In this paper, however, we will take a different
approach and `construct' a $q$-identity by repeatedly using the basic
identity (1.1) with the underlying resolution of $\bS(V)$ (section 2.3)
as a guiding principle.

Conjecturally, the resulting alternating sum formula will also arise
by applying the Euler-Poincar\'e principle to a certain resolution of
$\bS(\wh V)$ which, in some sense, is an appropriately `affinized'
version of Taylor's resolution of $\bS(V)$. \medskip

In the remainder of this section we will explain the algorithm and 
some of the properties of the resulting $q$-identity.  In section 
5 we will illustrate the algorithm in a few examples.

Consider the expression
$$
{q^{\sum_{(i,j) \in \cP} M_iM_j } \over \qn{M_1}\ldots\qn{M_n} }\,.
\tag{4.4}
$$
We now construct an alternating sum formula, bearing close resemblance
to the Taylor resolution, as follows
\item{$\bullet$}  Order the quadratic monomials $x_ix_j$, $(i,j)\in\cP$,
in any arbitrary way $\{ f_1=x_{i_1}x_{j_1},\ldots,f_t
=x_{i_t}x_{j_t}\}$.  Then, apply (1.1) to
the term 
$$
{q^{M_{i_1}M_{j_1}} \over \qn{M_{i_1}}\qn{M_{j_1}} }
$$
in (4.4), 
calling the summation variable $m_1$.  We proceed with the term in
(4.4) corresponding to $f_2$.  If $f_{12}= f_1f_2$ (cf.\ section 2.3),
we can apply (1.1) immediately.  On the other hand,
if $f_{12}\neq f_1f_2$ 
then one of the variables $M_{i_1}$ or $M_{j_1}$ appears in $M_{i_2}M_{j_2}$,
and the corresponding term in the denominator will have been shifted by
$m_1$.  Making the corresponding shift in the exponent, i.e.,
writing $MM' = (M-m_1)M' + m_1M'$, we can apply (1.1) to the
$(M-m_1)M'$ part, denoting the summation variable by $m_2$.  We continue
this process untill all terms $M_iM_j$, $(i,j)\in\cP$, in the 
exponent of (4.4) have been replaced.
The resulting expression will be an alternating 
sum with summation variables $m_k$, $k=1,\ldots,t$, in 1--1 correspondence
with the generating monomials $f_{k} = x_{i_k}x_{j_k}$.  The 
$M_i$ dependent remnant
in the $q$-exponent will be of the form
$$
\sum_{ I \in \cI_2\atop I = \{k\}\cup \{l\} } \ d_I\, m_k\, M_{j_l}\,,
\tag{4.5}
$$
where $d_I=0$ if $f_{kl} = f_kf_l$,
and $d_I=1$ if $f_{kl} \neq f_kf_l$.
\item{$\bullet$} In the second step we repeat the procedure to the
monomials $m_kM_{j_l}$, incorporating the appropriate shifts in the
$m_k$ and $M_{j_l}$, and calling the corresponding summation variables
$m_{kl}$.  Clearly, the summation variables introduced in this step
are in 1--1 correspondence with the $f_I$, $I\in\cI_2$, such that
$f_{kl}\neq f_kf_l$.  Note that, in (4.5), it can happen that $j_{l} =
j_{l'}$ for some $l\neq l'$ (cf.\ section 5.4 for an example).  In
that case it is important to keep the terms separate and remember
their origin.  The $M_i$ dependent remnant in the $q$-exponent will
now be of the form
$$
\sum_{ I \in \cI_3\atop I = \{k,l\}\cup \{m\} } \ d_I\, m_{kl} \, M_{j_m}\,,
\tag{4.6}
$$
where $d_I=0$ if $f_{klm} = f_{kl}f_m$ and $1$ otherwise.
\item{$\bullet$} Continue the procedure as before until all $M_i$
dependent parts in the $q$-exponent have been replaced by alternating sums.
\item{$\bullet$} As a last step we shift all the summation variables
$m_I$ such that they appear in the denominator as $\prod_I \qn{m_I}$.
\medskip

The resulting identity will be of the form 
$$
{q^{\sum_{(i,j) \in \cP} M_iM_j } \over \qn{M_1}\ldots\qn{M_n} } 
\eql  \sum_{m_I \geq0,\,I\in\cI'} 
(-1)^{\sum_{I\in\cI'} |I| m_I } 
{q^{Q(m_I)} \over \prod_{I\in\cI'} \qn{m_I} \prod_{i} \qn{M_i - 
\De M_i} }  \,,
\tag{4.7}
$$
where $\cI'$ is the subset of $\cI$ consisting of all sets
$\{i_1,\ldots,i_s\}$  such that 
$f_{i_1,\ldots,i_s} \neq f_{i_1,\ldots,\widehat{i_r},\ldots,i_s}
f_{i_r}$ for some $1\leq r\leq s$.
Furthermore,
$$
\De M_i  \eql \sum_{I\in\cI'}  a^{(i)}_I\, m_I \,,\tag{4.8}
$$
where the $a^{(i)}_I$ are a set of positive integers
such that $a^{(i)}_I\neq0$ iff $x_i$ occurs in $f_I$, and
$$
Q(m_I) \eql \half \sum_I\ |I|\, m_I(m_I-1) + Q'(m_I)\,,\tag{4.9}
$$
for some positive definite bilinear form $Q'$.  

Some more features of the expression (4.7) -- (4.9) can be derived by
examining how the Hilbert function of the underlying variety $V$ is
reproduced.  To this end it is convenient to multiply both 
sides by $\prod_i y_i^{M_i}$ and sum over $M_i\geq0$ (cf.\ the discussion
in section 1).
On the left hand side of (4.7), the $\cO(q^0)$-terms obviously
correspond to a basis for $\bS(V)$ (cf.\ remark 4.1).  On the right
hand side we get contributions only from $m_I=0$ or $m_I=1$.  If 
$m_I=0$ for all $I$ we find a contribution
$$
\prod_{i=1}^n\ {1\over (1-y_i)} \,, \tag{4.10}
$$
while if $m_I=1$, and $m_J=0$ for all $J\neq I$ contributes, 
the contribution will be
$$
(-1)^{|I|} \prod_{i=1}^n\ {y_i^{a_I^{(i)}} \over (1-y_i) } \,, \tag{4.11}
$$
to be compared to (2.10).  
If all $I\in \cI$ would contribute to the right hand side of
(4.7) through (4.11), then we would get exactly the expression (2.10)
corresponding to Taylor's resolution (cf.\ examples 5.1 and 5.4).  
However, the sum in (4.7) is over $I\in\cI'\subset \cI$ and in general 
$\cI' \neq \cI$.  The `missing terms' in (2.10) are recovered as follows.
Suppose $I=\{ i_1,\ldots,i_s\} = J \cup \{ i_r\}$ for some $1\leq r\leq s$,
and such that $f_I = f_J f_{i_r}$, 
i.e., $I\not\in\cI'$.  Then the positive definite bilinear form 
$Q'(m_I)$ in (4.9) will not contain a term 
$m_Jm_{i_r}$.  In other words,
the term in the summation on the right hand side of (4.7) with
$m_J=1$, $m_{i_r}=1$ and $m_I=0$,
for all other $I$, will contribute to the $\cO(q^0)$-term. The 
contribution is exactly (cf.\ example 5.2)
$$
(-1) (-1)^{|J|} \prod_{i=1}^n \ { y_i^{a_J^{(i)}+a_{i_r}^{(i)}} \over 
(1-y_i) } 
\eql (-1)^{|I|} \prod_{i=1}^n\ {y_i^{a_I^{(i)}} \over (1-y_i) } 
\,.\tag{4.12}
$$ 
This is one way the affinized expression (4.7) `knows about' Koszul
parts in the Taylor resolution (cf.\ remark 2.1)
and automatically takes care of them
without having to introduce, in some sense trivial, additional summation
variables.

It may also happen that the bilinear form $Q'(m_I)$ in (4.9) still
contains quadratic pieces $m_I^2$ for some $I$.  In that case the term
with $m_I=1$ and $m_J=0$, for all other $J\in\cI'$, will not
contribute to the $\cO(q^0)$-term on the right hand side of (4.7).
This will only happen if there exists another $I'\in\cI$, for which
the same thing happens and for which $f_I = f_{I'}$, $|I|=-|I'|$,
i.e., in that case the contributions from $I$ and $I'$ in Taylor's
resolution will cancel (cf.\ example 5.4).  Whenever this happens it
might indicate that Taylor's resolution is not a minimal resolution
and that it can be reduced by removing the spaces corresponding to $I$
and $I'$ (cf.\ the example in section 2.3).

If in Taylor's resolution there exists an $I\in\cI$ such that $f_I =
f_J f_{J'}$ for some $J \cup J' \subset I$, this is another indication
that the resolution might not be a minimal one.  In that case, in the
affinized expression (4.7), it might be possible to explicitly sum out
the summation variable $m_I$ to obtain an expression associated to
some reduction of Taylor's resolution (cf.\ example 5.2).  This is
another way in which the affinized expression (4.7) knows about Koszul
parts in Taylor's resolution.

Finally, it should be obvious that the final form of the identity
(4.7) is not necessarily unique but could depend on the order in which
the various summation variables $m_I$ are introduced.  In principle we
could fix the expression by specifying the order of $m_I$ (e.g.,
through a reverse graded lexicographic ordering on the $I$), but in
practise the identities are more easily accessible by using already
established identities for sub-ideals as `building blocks' (cf.\
examples 5.3 and 5.5).  Also, it might very well be that specific
identities are more `manageable' or `useful' than others.

\head 5. Examples \endhead

In this section we will illustrate the algorithm outlined in section 4.2
and some of the properties of the resulting $q$-identities 
by explicitly going through a few examples of quadratic monomial
ideals.  The main results are the identities (5.2), (5.6), (5.10), (5.13),
(5.19) and  (5.23).
The results can be used as building blocks for more complicated 
examples.

\subhead 5.1. $I = \langle x_1x_2,x_2x_3\rangle$ \endsubhead\shskip

Consider the ideal $I = \langle x_1x_2,x_2x_3\rangle\subset 
\CC[x_1,x_2,x_3]$.  The corresponding variety $\bV(I)$ is a 1-dimensional
subvariety of $\PP^2$ (union of a line and a point).
Taylor's resolution (2.17) of $\bS(V)$ takes the form
$$
0 @>>> \bS(-1,-1,-1) @>>> \bS(-1,-1,0) \oplus \bS(0,-1,-1) @>>>
\bS @>>>\bS(V) @>>> 0 \,. \tag{5.1}
$$
By repeated application of the basic identity (1.1), following the 
algorithm outlined in section 4.2, we find
$$ \align
& { q^{M_1M_2+M_2M_3} \over \qn{M_1}\qn{M_2}\qn{M_3} } \cr & \eql
\sum_{m_1} (-1)^{m_1} {q^{ {1\over2} m_1(m_1-1)} \over \qn{m_1}}
{q^{M_2M_3} \over \qn{M_1-m_1}\qn{M_2-m_1}\qn{M_3} } \\
& \eql
\sum_{m_1} (-1)^{m_1} {q^{ {1\over2} m_1(m_1-1)} \over \qn{m_1}}
{q^{(M_2-m_1)M_3 + m_1M_3} \over \qn{M_1-m_1}\qn{M_2-m_1}\qn{M_3} } \\
& \eql
\sum_{m_1,m_2} (-1)^{m_1+m_2} {q^{ {1\over2} m_1(m_1-1)+{1\over2} m_1(m_1-1)} 
\over \qn{m_1}\qn{m_2}} {q^{m_1M_3} 
\over \qn{M_1-m_1}\qn{M_2-(m_1+m_2)}\qn{M_3-m_2} } \\
& \eql
\sum_{m_1,m_2} (-1)^{m_1+m_2} {q^{ {1\over2} m_1(m_1-1)+{1\over2} m_1(m_1-1)} 
\over \qn{m_1}\qn{m_2}} {q^{m_1(M_3-m_2)+m_1m_2} 
\over \qn{M_1-m_1}\qn{M_2-(m_1+m_2)}\qn{M_3-m_2} } \\
& \eql\sum_{m_1,m_2,m_{12}} 
(-1)^{m_1+m_2+m_{12}} {q^{ {1\over2} m_1(m_1-1)+{1\over2} m_1(m_1-1)
+{1\over2} m_{12}(m_{12}-1)} 
\over \qn{m_1-m_{12}}\qn{m_2}\qn{m_{12}}} \\
& \qquad\qquad \times {q^{m_1m_2} 
\over \qn{M_1-m_1}\qn{M_2-(m_1+m_2)}\qn{M_3-(m_2+m_{12})} } \\
& \eql
\sum_{m_1,m_2,m_{12}} 
(-1)^{m_1+m_2} {q^{ {1\over2} m_1(m_1-1)+{1\over2} m_2(m_2-1)
+m_{12}(m_{12}-1)} 
\over \qn{m_1}\qn{m_2}\qn{m_{12}}} \\
& \qquad\qquad \times 
{q^{m_1m_2 + m_{12}(m_1+m_2)} 
\over \qn{M_1-(m_1+m_{12})}
\qn{M_2-(m_1+m_2+m_{12})}\qn{M_3-(m_2+m_{12})} } \,, \tag{5.2}\endalign
$$
where, in the last step, we have shifted the summation variable 
$m_1\to m_1 + m_{12}$.
\bigskip
Indeed, (5.2) is of the form (4.7) with $\cI'=\cI=\{1,2,12\}$,
$$ 
\De M_1\eql m_1+m_{12}\,,\quad 
\De M_2\eql m_1+m_2+m_{12} \,,\quad
\De M_3\eql m_2+m_{12}\,,\tag{5.3}
$$
and 
$$
Q'(m_I) \eql m_1m_2 + m_{12}(m_1+m_2)\,.\tag{5.4}
$$
The $\cO(q^0)$-term in the resulting identity for the Hilbert series 
$h_{\wh V}(y;q)$ leads to the identity
$$
{ 1\over (1-y_1)(1-y_3)} + {y_2\over 1-y_2} \eql
{ 1 - y_1y_2 - y_2y_3 + y_1y_2y_3 \over (1-y_1)(1-y_2)(1-y_3)} \,.\tag{5.5}
$$

\subhead 5.2. $I= \langle x_1x_2,x_2x_3,x_3x_4 \rangle$ \endsubhead\shskip

Consider the ideal $I = \langle x_1x_2,x_2x_3,x_3x_4\rangle\subset 
\CC[x_1,x_2,x_3,x_4]$. The resolution of $\bS(V)$ was discussed in 
(2.19).  Applying (5.2) we find 
$$ \align
& { q^{M_1M_2+M_2M_3+M_3M_4} \over \qn{M_1}\qn{M_2}\qn{M_3}\qn{M_4} } \\
& \eql
\sum_{m_I\geq0\atop I=1,2,12} 
(-1)^{m_1+m_2} {q^Q 
\over \prod_{I=1,2,12} \qn{m_I}
\prod_{i=1}^3\qn{M_i-\De M_i}  } 
{q^{M_3M_4} \over \qn{M_4} }\,,
\endalign
$$
where $\De M_i$, $i=1,2,3$, is given by (5.3) and $Q$ by (4.9) and (5.4).
Now write 
$$
M_3M_4 \eql (M_3-(m_2+m_{12}))M_4 + (m_2+m_{12})M_4\,,
$$
and apply (1.1) to the first term with summation variable $m_3$.
Then write, in the $q$-exponent,
$$
(m_2+m_{12})M_4 \eql (m_2+m_{12})(M_4-m_3) + (m_2+m_{12})m_3\,,
$$
and apply (1.1) to $m_{12}(M_4-m_3)$ with summation variable $m_{123}$.
Finally, writing
$$
m_2(M_4-m_3) \eql m_2(M_4-(m_3+m_{123})) + m_2 m_{123}\,,
$$
and applying (1.1) to $m_2(M_4-(m_3+m_{123}))$ with summation variable 
$m_{23}$ and shifting $m_2\to m_2 + m_{23}$ and $m_{12}\to m_{12} + m_{123}$,
yields
$$
{ q^{M_1M_2+M_2M_3+M_3M_4} \over \qn{M_1}\qn{M_2}\qn{M_3}\qn{M_4} } 
\eql
\sum_{m_I\geq0 \atop I\in\cI'} (-1)^{\sum_I |I| m_I}
{q^{Q(m_I)} \over \prod_{I\in \cI'} \qn{m_I} 
\prod_{i=1}^4 \qn{M_i-\De M_i} } \,,\tag{5.6}
$$
where $\cI' = \{ 1,2,3,12,23,123\}$, 
$$ \align
\De M_1 & \eql m_1+m_{12}+m_{123}\,,\\
\De M_2 & \eql m_1+m_2+m_{12}+m_{23}+m_{123}\,,\\
\De M_3 & \eql m_2+m_3+m_{12}+m_{23}+m_{123}\,,\\
\De M_4 & \eql m_3+m_{23}+m_{123}\,,\tag{5.7}
\endalign
$$
and
$$ \align
Q \eql & \half \sum_{I\in\cI'} |I| m_I (m_I-1) 
+ m_1m_2 + m_2m_3 + m_{12}m_{23} \\ 
& + (m_1+m_2+m_3)(m_{12}+m_{23}+m_{123}) 
 + 2(m_{12}+m_{23})m_{123} \,.\tag{5.8}\endalign
$$
Observe, indeed, that since $f_{13}=f_1f_3$, the subset $\{1,3\}$ is absent 
from $\cI'$ and hence the corresponding summation variable $m_{13}$ does
not occur in (5.6).

Moreover, as discussed in section 2.3, 
the Taylor resolution (2.19) of $\bS(V)$ is not minimal.  A minimal
resolution is obtained from (2.19) by removing the spaces corresponding
to $I=13$ and $I=123$.  This manifests itself in (5.6) by the fact that
the summation variable $m_{123}$ can be summed out.

First, notice that we can get rid of the $m_{123}$ shifts in the
$q$-numbers in the denominator, by shifting 
$m_1 \to m_1 - m_{123}$ and $m_3\to m_3-m_{123}$.  This yields an
exponent
$$ 
Q ~\to~ Q' + \half m_{123}(m_{123}-1) \,, 
$$
with 
$$\align
Q' \eql & \half \sum_{i=1,2,3} m_i(m_i-1) + \sum_{ij=12,23} m_{ij}(m_{ij}-1)
\\ & + m_1m_2 + m_2m_3 + (m_1+m_2+m_3)(m_{12}+m_{23}) +  
  m_{12}m_{23}\,.\tag{5.9}\endalign
$$
Next, we can sum out $m_{123}$ by (1.1) after which we obtain
$$
{ q^{M_1M_2+M_2M_3+M_3M_4} \over \qn{M_1}\qn{M_2}\qn{M_3}\qn{M_4} } \eql
\sum_{m_I\geq0\atop I\in\cI''} (-1)^{\sum_I |I|m_I}
{q^{Q''} \over \prod_{I\in\cI''} \qn{m_I} \prod_{i=1}^4 \qn{M_i-\De M_i}}\,,
\tag{5.10}
$$
where $\cI''=\{1,2,3,12,23\}$,
$$ \align
\De M_1 & \eql m_1+m_{12}\,,\\
\De M_2 & \eql m_1+m_2+m_{12}+m_{23}\,,\\
\De M_3 & \eql m_2+m_3+m_{12}+m_{23}\,,\\
\De M_4 & \eql m_3+m_{23}\,,\tag{5.11}
\endalign
$$
and
$$
Q'' \eql Q' + m_1m_3\,,\tag{5.12}
$$
with $Q'$ given by (5.9).

\subhead 5.3. $I=\langle x_1x_2,x_2x_3,\ldots,x_{n-1}x_n\rangle$
\endsubhead\shskip

The ideal $I=\langle x_1x_2,x_2x_3,\ldots,x_{n-1}x_n
\rangle \subset \CC[x_1,\ldots,x_n]$ corresponds to a dimension
$\left[ {n-1\over2}\right]$ variety $V_n\subset\PP^{n-1}$ and generalizes
the examples of sections 3.2, 5.1 and 5.2.
The corresponding $q$-identity can be proved by induction. The result is 
$$
{q^{\sum_{i=1}^{n-1} M_iM_{i+1} } \over \qn{M_1}\ldots\qn{M_n} } \eql
\sum_{m_1,\ldots,m_{n-1}\atop n_1,\ldots,n_{n-2} }
(-1)^{\sum m_i} {q^{Q(m_i,n_i)} \over \prod_i \qn{m_i}\qn{n_i} \prod_{i=1}^n 
\qn{M_i - \De M_i} }\,, \tag{5.13}
$$
where 
$$\align
Q \eql & \half \sum_{i=1}^{n-1} m_i(m_i-1) + 
\sum_{i=1}^{n-2} n_{i} (n_{i}-1) + 
\sum_i m_i (m_{i+1}+m_{i+2}) \\ & + \sum_{i=1}^{n-1} 
m_i (n_{i-2} + n_{i-1} + n_{i} + n_{i+1} + n_{i+2} )
+ \sum_{i=1}^{n-2} n_{i} (n_{i+1} + n_{i+2}) \,,\tag{5.14}\endalign
$$
and 
$$
\De M_i \eql m_i + m_{i-1} + n_i + n_{i-1}+n_{i-2} \,.\tag{5.15}
$$
For simplicitly of notation we have denoted $n_i=m_{i\, i+1}$ and
$m_n\equiv m_0 \equiv n_0 \equiv n_{-1} \equiv n_{n-1} \equiv n_n \equiv 0$.
Note that, for $M_j=0$, the $q$-identity factorizes and reduces to 
the same identity for smaller $n$. 

The induction procedure that leads to (5.13) suggests the following
recursion relation for the Hilbert series $h_n(y)$ of the underlying
variety $V_n$
$$
h_n(y_1,\ldots,y_n) \eql {1\over 1-y_n} h_{n-2}(y_1,\ldots,y_{n-2}) +
  {y_{n-1}\over 1-y_{n-1} } h_{n-3}(y_1,\ldots,y_{n-3}) \,,\tag{5.17}
$$
with $h_{0}= 1$, $h_1={1\over 1-y_1}$ and 
$h_2 = {1\over 1-y_2} + {y_1\over 1-y_1}$.

\subhead 5.4. $I=\langle x_1x_2,x_2x_3,x_1x_3\rangle$ \endsubhead\shskip

To obtain a $q$-identity for the ideal $I=\langle x_1x_2,x_2x_3,x_1x_3
\rangle \subset \CC[x_1,x_2,x_3]$, corresponding to three non-collinear
points in $\PP^2$ (cf.\ \cite{Ha, Example 13.11}),
we need the following lemma
$$
{q^{2MN} \over \qn{M}\qn{N} } \eql \sum_{r,s,t}(-1)^{r+s}
{q^{ {1\over2} r(r-1)+ {1\over2} s(s-1) + t(t-1) +
  (rs+rt+st) + (r+t)M} \over
  \qn{r}\qn{s}\qn{t} \qn{M-(r+s+t)}\qn{N-(r+s+2t)} } \,,\tag{5.18}
$$
which is proved by the same techniques as before, i.e.,
write $2MN = MN + MN$ and apply (1.1) to $MN$ with summation 
variable $r$.  Then in the remaining exponent
write $MN = (M-r)(N-r) +rN + rM -r^2$.
Apply (1.1) again, now to $(M-r)(N-r)$ with summation variable $s$
and write in the remaining exponent $rN = r(N-(r+s)) + r(r+s)$ and apply
(1.1) to $r(N-(r+s))$ with summation variable $t$.  Finally, shift
$r\to r+t$.  This yields (5.18).\medskip

Now, in 
$$
{ q^{M_1M_2+M_2M_3+M_1M_3} \over \qn{M_1}\qn{M_2}\qn{M_3}}\,,
$$
we apply (1.1) consecutively to
$M_1M_2$, $(M_2-m_1)M_3$, $(M_1-m_1)(M_3-m_2)$ and $m_2M_1$ with summation 
variables $m_1,m_2,m_3$ and $m_{23}$.  This yields
$$ \align
\sum_{m_I\geq0\atop I=1,2,3,23} & (-1)^{\sum m_I} 
{q^Q \over \qn{m_1} \qn{m_2-m_{23}} \qn{m_3} \qn{m_{23}} }  \\ & \times 
{1\over
\qn{M_1-(m_1+m_3+m_{23})} \qn{M_2-(m_1+m_2)} \qn{M_3-(m_2+m_3)} }\,,
\endalign
$$
with 
$$
Q \eql \half \sum_{i=1,2,3} m_i(m_i-1) + \half m_{23}(m_{23}-1) + m_2m_3
+ 2 m_1 M_3\,.
$$
Then write $2m_1M_3 = 2m_1(M_3-(m_2+m_3))$ and apply (5.18) with
the substitutions $r\to m_{13}$, $s\to m_{12}$ and $t\to m_{123}$.  In the
result shift $m_1\to m_1 + m_{12}+m_{13}+m_{123}$ and $m_2 \to 
m_2 + m_{23}$.  This finally yields 
$$ 
{ q^{M_1M_2+M_2M_3+M_1M_3} \over \qn{M_1}\qn{M_2}\qn{M_3}} 
\eql \sum_{m_I\geq0\atop I\in\cI'} (-1)^{\sum_I |I|m_I} 
{q^{Q(m_I)} \over \prod_{I\in\cI'} \qn{m_I} 
\prod_{i=1}^3 \qn{M_i-\De M_i } }\,,\tag{5.19}
$$
with $\cI'= \{1,2,3,12,23,13,123\} = \cI$, 
$$\align
\De M_1 & \eql m_1+m_3+m_{12}+ m_{23}+ m_{13}+m_{123} \,,\\
\De M_2 & \eql m_1+m_2 + m_{12}+ m_{23}+ m_{13}+m_{123}\,, \\
\De M_3 & \eql m_2+m_3+ m_{12}+ m_{23}+ m_{13}+2m_{123}\,,\tag{5.20} 
\endalign
$$
and
$$ \align
Q \eql & \half \sum_{I\in\cI'} |I| m_I (m_I-1) 
+ m_{13}^2 + m_{123}^2 \\ & 
+ 2m_1(m_2+m_3) + m_2m_3 + 2 m_{23} (m_{12}+m_{13}) + 3m_{12}m_{13} \\
& + m_{12}(m_1+2m_2+2m_3) + m_{23}(2m_1+m_2+m_3) + m_{13} (m_1+m_2+m_3) \\
& + 2m_{123} (m_1+m_2+m_3) + m_{123}(3m_{12} + 2 m_{23} + 4m_{13}) \,.
\tag{5.21} \endalign
$$
The $\cO(q^0)$-term in the resulting identity for the Hilbert series 
$h_{\wh V}(y;q)$ leads to the identity
$$
{1\over 1-y_1} + {y_2\over 1-y_2} + {y_3\over 1-y_3} 
\eql { 1- y_1y_2 - y_2y_3 - y_1y_3 + 2 y_1y_2y_3 \over (1-y_1)
(1-y_2)(1-y_3) } \,. \tag{5.22}
$$
Note that in deriving (5.22) the terms $m_I=1$ and all others vanishing,
do not contribute for $I=13$ and $I=123$ due to the terms $m_I^2$ in 
(5.21).  Indeed, since $f_{13}=f_{123}$ their contribution to 
$h_{V}(y)$ in (2.10) would cancel.  This is related to the fact that,
also in this case, Taylor's resolution is not minimal but can be 
further reduced by omitting the spaces corresponding to $I=13$ and $I=123$.
So, even though it does not seem possible to further simplify the
affine identity (5.19), the identity somehow knows about the non-minimality
of Taylor's resolution.

\subhead 5.5. $I=\langle x_1x_2,x_2x_3,\ldots,x_{n-1}x_n, x_1x_n\rangle$
\endsubhead\shskip

As a generalization of the example in section 5.4, consider the 
ideal \hfil\break
$I=\langle x_1x_2, x_2x_3, \ldots, x_{n-1}x_n, x_1x_n\rangle \subset
\CC[x_1,\ldots,x_n]$, $n\geq6$.  Using, as an intermediate
step, the result (5.13) we staightforwardly find
$$
{q^{M_1M_2+\ldots M_{n-1}M_n + M_1M_n} \over \prod_{i=1}^n \qn{M_i} }\eql
\sum_{m_1,\ldots,m_n\atop {n_1,\ldots,n_n \atop p,p'}}\
(-1)^{\sum m_i + p + p'} { q^Q \over 
\prod_{i} \qn{m_i}\qn{n_i} \qn{p} \qn{p'}\qn{M_i - \De M_i} }\,,\tag{5.23}
$$
with \footnote{For $n=4,5$ the relevant formulae are the obvious 
concatenation of these.}
$$ \align
\De M_1 & \eql m_1+ m_n + n_1 + n_n + n_{n-1} + p + p' \,,\\
\De M_2 & \eql m_2+ m_1 + n_2 + n_1 + n_n + p \,,\\
\De M_3 & \eql m_3+ m_2 + n_3 + n_2 + n_1 + p \,,\\
\De M_i & \eql m_i + m_{i-1} + n_i + n_{i-1} + n_{i-2} \,,\qquad 
   4\leq i \leq n-3\,,\\
\De M_{n-2} & \eql m_{n-2} + m_{n-3} + n_{n-2} + n_{n-3} + n_{n-4} + p' \,,\\
\De M_{n-1} & \eql m_{n-1} + m_{n-2} + n_{n-1} + n_{n-2} + n_{n-3} + p' \,,\\
\De M_n     & \eql m_n+ m_{n-1} + n_n + n_{n-1} + n_{n-2} + p + p' \,,
\tag{5.24}\endalign
$$
where the subscripts on $m_i$ and $n_i$ have to be taken modulo $n$.
In addition to the notation in section 5.3, 
we have denoted $n_{n-1}=m_{n-1\, n}$, 
$n_n = m_{1\, n}$, $p = m_{1\,2\,n}$ and $p'=m_{n-2\, n-1\,n}$.
The explicit expression for $Q(m_i,n_i,p,p')$ in (5.23) is left as 
an exercise to the reader.

\head 6. Concluding remarks \endhead

In this paper we have explained an algorithm to associate a
$q$-identity to an arbitrary projective variety $V$ defined by a
quadratic monomial ideal.  The identities were argued to correspond to
two different ways of computing the Hilbert series of a suitable
`affinization' $\wh V$ of the variety $V$, on the one hand by
computing an explicit basis for the coordinate ring $\bS(\wh V)$, on
the other hand by constructing a free resolution of this coordinate
ring.  The algorithm was illustrated in numerous examples.

The algorithm is based on Taylor's resolution for the coordinate ring
$\bS(V)$ of the underlying (finite-dimensional) projective variety
$V$.  This resolution is not always a minimal free resolution.  In
section 5.2 we have seen an example where the identity can be further
reduced to an identity which one would like to associate with the 
minimal resolution in that case.  This further reduction can typically
be done for the `Koszul parts' in Taylor's resolution.  In other cases,
such as in the example of section 5.4, a further reduction does not appear 
to be possible even though Taylor's resolution is not minimal.   In that 
example, i.e., $I=\langle x_1x_2,x_2x_3,x_1x_3\rangle \subset
\CC[x_1,x_2,x_3]$, the minimal resolution of $\bS(V)$ looks like 
$$ 
0 @>>>  \bS(-3)^2 @>>> \bS(-2)^3 @>>> \bS @>>> \bS(V) @>>> 0\,.
\tag{6.1}
$$
[Here we have only indicated the shift in total degree.]
Surprisingly, however, one can find a $q$-identity which one 
would like to associate to (6.1).  It reads
$$
{ q^{M_1M_3} \over \qn{M_1}\qn{M_2}\qn{M_3} } \eql
\sum_{m_I\geq0\atop I=1,2,3,12,23}\ (-1)^{\sum_i m_i} 
{q^{Q(m_I)} \over \prod_I \qn{m_I} } {1\over \prod_i \qn{M_i-\De M_i} }\,,
\tag{6.2}
$$
where 
$$ \align
\De M_1 & \eql m_1+m_2+m_{12} \,,\\
\De M_2 & \eql m_1+m_3+m_{12}+m_{23} \,,\\
\De M_3 & \eql m_2+m_3+m_{23} \,,
\tag{6.3}\endalign
$$
and
$$\align
Q(m_I) \eql & \half \sum_{i=1,2,3} m_i(m_i-1) +
\sum_{I=12,23} m_I(m_I-1) + m_1m_2+m_1m_3+m_2m_3 \\
& + (m_1+m_2+m_3)(m_{12}+m_{23}) + m_{12}m_{23}\,. 
\tag{6.4}\endalign
$$
The relation of (6.2) to the coordinate ring of $\bS(\wh V)$ is however
not clear to me at present.

Even though we have restricted our attention to varieties defined by
quadratic monomial ideals, the idea is far more general.  Indeed, one
can often find interesting identities associated to more general ideals.
Consider, e.g., the `trivial' example of $I=\langle x\rangle \subset
\CC[x]$.  Obviously, both $\bS(V)$ and $\bS(\wh V)$ only contain the
constant polynomials.  The resolution of  $\bS(\wh V)$ gives rise,
however, to the not completely trivial (but well-known) $q$-identity
$$
\sum_{m\geq0} \ (-1)^m \ {q^{{1\over2}m(m-1)} \over \qn{m} \qn{M-m}} \eql
\de_{M,0}\,. 
\tag{6.5}
$$

For a slightly less trivial example, consider the ideal
$I=\langle x_1, x_2x_3(x_2-x_3) \rangle \subset \CC[x_1,x_2,x_3]$
corresponding to complete intersection of a line and a cubic, i.e.,
three collinear points in $\PP^2$ (cf.\ \cite{Ha, Example 13.11}).
The resolution of $\bS(V)$ is Koszul's resolution
$$
0 @>>> \bS(-4) @>>> \bS(-1) \oplus \bS(-3) @>>> \bS @>>> \bS(V) @>>> 0\,,
\tag{6.6}
$$
and the associated $q$-identity is 
$$
\sum_{m_I\geq0\atop I=1,2,12}\ (-1)^{m_1+m_2} 
{q^{Q(m_I)} \over \prod_I \qn{m_I} } {1\over \prod \qn{M_i-\De M_i} }
\eql \de_{M_1,0} \sum_{m\geq0} (-1)^m {q^{{1\over2}m(m-1)}\over \qn{m}
\qn{M_2-2m}\qn{M_3-m} } \,,
\tag{6.7}
$$
where
$$
\De M_1 \eql m_1+m_{12} \,,\quad
\De M_2 \eql 2m_2 + 2m_{12} \,,\quad
\De M_3 \eql m_2+m_{12}\,,\tag{6.8}
$$
and
$$
Q\eql \half \sum_{i=,1,2} m_i(m_i-1) + m_{12}(m_{12}-1) + m_1m_2
  + (m_1+m_2) m_{12} \,.\tag{6.9}
$$
Equation (6.7) can be proved by shifting $m_2\to m_2-m_{12}$, summing
over $m_{12}$ by (1.1) and then performing the sum over $m_1$ by
(6.5).

Despite the existence of examples of $q$-identities for other than
quadratic monomial ideals, we believe the ones corresponding to
quadratic monomial ideals are `the nicests' and are the ones most
relevant for the application in conformal field theory.  In a sequel
to this paper we discuss $q$-identities associated to flag
varieties \cite{BH}.  The corresponding Hilbert series correspond to the
partition functions of quasi-particles in WZW conformal field theories
and are the building blocks for characters of affine Lie algebras.  In
fact, we will argue that the partial Hilbert series of an affinized flag
variety is, upto a trivial factor, precisely the modified
Hall-Littlewood polynomial.

Flag varieties are defined by an ideal of (non-monomial) quadratic relations.
Nevertheless, we will show that, at least as far as the computation 
of the Hilbert series is concerned,  the computations can be reduced 
to those for quadratic monomial ideals discussed in this paper.

\head Acknowledgements \endhead

I would like to thank Omar Foda, Emily Hackett-Jones and David Ridout
for discussions.  P.B.\ is supported by a \qeii\ research fellowship
from the Australian Research Council.



\Refs
\widestnumber\key{DKKMM}

\ref \key An
\by G.E.~Andrews
\book The theory of partitions 
\bookinfo Encycl.\ of Math.\ and its Appl.\ vol.\ {\bf 2}
\publ Addison-Wesley
\publaddr Reading
\yr 1976
\endref

\ref \key BH
\by P.~Bouwknegt and N.~Halmagyi
\paper $q$-identities and affinized projective varieties, II: Flag varieties
\paperinfo ADP-99-2/M77, in preparation
\endref


\ref \key BS1
\by P.~Bouwknegt and K.~Schoutens
\paper The $\widehat{SU(n)}_1$ WZW models: Spinon decomposition and
       Yangian structure
\jour Nucl.\ Phys.\
\vol B482 \yr 1996 \pages 345--372
\finalinfo [{\tt hep-th/9607064}]
\endref


\ref \key BS2
\bysame
\paper Spinon decomposition and Yangian structure of 
       $\widehat{\frak{sl}_n}$ modules
\inbook Geometric Analysis and Lie Theory in Mathematics and
        Physics, Australian Mathematical Society Lecture Series {\bf 11}
\eds A.L.~Carey and M.K.~Murray 
\publ Cambridge University Press
\publaddr Cambridge
\yr 1997 \pages 105--131
\finalinfo [{\tt q-alg/9703021}]
\endref

\ref \key CLO1
\by D.~Cox, J.~Little and D.~O'Shea
\book Ideals, varieties, and algorithms: An introduction to computational 
      algebraic geometry and commutative algebra
\publ Springer Verlag
\publaddr  Berlin \yr 1997
\endref

\ref \key CLO2
\bysame
\book Using algebraic geometry 
\publ Springer Verlag
\publaddr  Berlin \yr 1998
\endref

\ref \key DKKMM
\by S.~Dasmahapatra, R.~Kedem, T.~Klassen, B.~McCoy and E.~Melzer
\paper Quasi-particles, conformal field theory and $q$-series
\jour Int.\ J.\ Mod.\ Phys.\  \vol B7 \yr 1993 \pages 3617--3648
\finalinfo [{\tt hep-th/9303013}]
\endref

\ref \key Ei
\by D.~Eisenbud
\book Commutative algebra; with view toward algebraic geometry
\bookinfo Graduate Texts in Math.\ 150
\publ Springer Verlag
\publaddr Berlin \yr 1994
\endref

\ref \key FS
\by B.L.~Feigin and A.V.~Stoyanovsky
\paper Quasi-particles models for the representations of Lie algebras
       and geometry of flag manifold 
\finalinfo [{\tt hep-th/9308079}]
\endref

\ref \key Ha
\by J.~Harris
\book Algebraic geometry: A first course
\publ Springer Verlag 
\publaddr Berlin
\yr 1992
\endref

\ref \key Ka
\by V.G.~Kac
\book Infinite dimensional Lie algebras 
\publ Cambridge University Press
\publaddr Cambridge \yr 1995
\endref

\ref \key KMM
\by R.~Kedem, B.~McCoy and E.~Melzer
\paper The sums of Rogers, Schur and Ramanujan and the Bose-Fermi
       correspondence in $1+1$-dimensional quantum field theory
\inbook in ``Recent progress in Statistical Mechanics and Quantum 
        Field Theory''
\eds P.~Bouwknegt et al.
\publ World Scientific \publaddr Singapore \yr 1995 \pages 195--219
\finalinfo [{\tt hep-th/9304056}]
\endref

\endRefs
\enddocument